# Classification revisited: a web of knowledge

Aida Slavic

UDC Consortium

## Abstract

*The vision of the Semantic Web (SW) is gradually unfolding and taking shape through a web of linked data, a part of which is built by capturing semantics stored in existing knowledge organization systems (KOS), subject metadata and resource metadata. The content of vast bibliographic collections is currently categorized by some widely used bibliographic classification and we may soon see them being mined for information and linked in a meaningful way across the Web. Bibliographic classifications are designed for knowledge mediation which offers both a rich terminology and different ways in which concepts can be categorized and related to each other in the universe of knowledge. From 1990-2010 they have been used in various resource discovery services on the Web and continue to be used to support information integration in a number of international digital library projects. In this chapter we will revisit some of the ways in which universal classifications, as language independent concept schemes, can assist humans and computers in structuring and presenting information and formulating queries. Most importantly, we highlight issues important to understanding bibliographic classifications, both in terms of their unused potential and technical limitations.*

## 1. Background

Classifications are created using the intellectual power of ontological observation and analysis and reflect whichever approach one decides to take in defining and grouping things. Whatever the motive for their creation, classifications are logically and semantically organized schemes of concepts representing some kind of reality and as such are essential and versatile tools for the representation and visualisation of a knowledge space. A classification, by nature, cannot serve all purposes, however, if it is universal with respect to knowledge coverage and structured in a comprehensive, detailed, highly elaborate and logical way, it is likely to be more versatile than a scheme created for one single subject, system or task.

Bibliographic or documentary classifications are a special kind of knowledge classification designed for the mediation of knowledge. Their unique feature is that they are not concerned with objects or entities, as other knowledge classifications, but rather with subjects, i.e. the ways in which entities are described in documents. Bibliographic classifications will systematise phenomena and topics that can be studied in relation to these phenomena, but will also provide the vocabulary necessary to denote types of knowledge presentation, points of view, the targeted audience or form of document. As subject is a complex construct which ought to be described with a series of concepts in various relationships, classification can also provide rules on how to express complex interactions between phenomena of knowledge as they appear in documents. The oldest and the best known type of bibliographic classifications are library classifications which are created primarily for the physical arrangement of library shelves. The fact that a classification is created for shelf arrangement, and not for detailed indexing for the purposes of metadata-based information retrieval, may influence its structure and its vocabulary as will be explained in the following sections. But whatever their original purpose, general bibliographic classifications are known to be complex systems.

Although there are many classifications, the three most internationally used general bibliographic classifications are Dewey Decimal Classification (DDC), Universal Decimal Classification (UDC) and Library of Congress Classification (LCC). They are used in the greatest number of countries and bibliographic collections and are considered *de facto* standards in information exchange. The main reason that these classifications prevail is a complex set of circumstances primarily related to the power of their ownership, services based on them and their continuous maintenance and development. DDC, although originally created for the shelf arrangement of a single college library, started to be widely used in Anglo-American libraries from 19th century onwards, as the best classification for shelf arrangement at the time. Its use throughout and beyond 20th century was promoted by the OCLC bibliographic service. UDC, originally created as an indexing language for an international universal bibliography project in 1895, gained widespread use through the international presence of its previous owner, the International Federation of Information (FID), which supported its development, translation into more than thirty languages and its world-wide distribution. LCC popularity can be attributed to its governmentally controlled and well supported administration, the speed with which works are classified and the availability of LCC classmarks on Library of Congress catalogue cards, rather than its inherent structural quality (Marcella & Newton, 1995: 60). Circumstances in the bibliographic domain are such that Colon Classification (CC), which is certainly the most theoretically praised library classification, is barely used, while Bliss Bibliographic Classification (BC2), which is supposed to demonstrate an excellence in classification design, is only 60% complete and is also barely used. There is also a number of national general classification systems but their use is confined to a single country or language which limits their role in the context of global information integration.

Intellectual indexing and classification requires a high level of expertise and is an expensive and time consuming process which may not be suited to, or required for, all information retrieval scenarios. For most text-retrieval tasks, various models of automatic text processing and advanced techniques such as statistical models, language models or machine learning, will perform well. However, not all documents are available in digital form and not all digital documents are textual, in the same language or the same script. Equally, not all digital collections are available or accessible in an open networked environment and will not lend themselves easily to processing by otherwise successful data and knowledge mining methods. The task of merging and integrating information contained in legacy collections into, what may be, a web of knowledge, is still ahead of us.

National bibliographies of most countries continuously collect, describe and classify everything that is published in their respective countries, by their citizens or in their official languages. This practice has existed for over a century, in some instances even longer, and has led to the creation of large bibliographic collections, only part of which may be available in digital form. Although this may hardly be comparable with the scale of information available on the Web, we are still talking about hundreds and hundreds of millions of documents published in numerous languages all around the world from the beginning of literacy to date. We can assume that in the foreseeable future libraries will continue to classify books for the purpose of collection management. and we know that legacy bibliographic data is something we will try to preserve.

The fact that document collections world-wide are classified using bibliographic classification systems may play a significant role in enabling subject searching across these collections. A significant increase in the use of classification in various information integration and discovery services from 1990-2010 clearly indicated such a trend. In the following section we will highlight some shared structural features of bibliographic classifications that are of particular importance for their use in information retrieval and discovery.

## 2. Classification, how does it work?
In Figure 1 below, we see two ways of representing a knowledge field for the purpose of knowledge browsing: one system using natural language terms and the other employing a systematic or classificatory arrangement

| | | |
|---|---|---|
| Antineutrinos | 539.1 | Nuclear physics. Atomic physics. **Molecular** physics |
| Antineutrons | 539.12 | Elementary and simple particles |
| Antiprotons | 539.123/.124 | Leptons. Including: Muons |
| Atomic physics | 539.123 | Neutrinos |
| Baryons | 539.123.6 | Antineutrinos |
| Beta-particles | 539.124 | Electrons (including **beta-particles**) |
| Bosons | 539.124.6 | **Positrons** |
| Electrons | 539.125/.126 | Hadrons. Baryons and mesons |
| Hadrons | 539.125 | Nucleons |
| Hyperons | 539.125.4 | Protons |
| Leptons | 539.125.46 | Antiprotons |
| Mesons | 539.125.5 | Neutrons |
| Mesons | 539.125.56 | Antineutrons |
| Molecular physics | 539.126.3 | Mesons |
| Muons | 539.126.4 | **Resonances** |
| Neutrinos | 539.126.6 | Hyperons |
| Neutrons | | |
| Nuclear physics | | |
| Nuclei | | |
| Nucleons | | |
| Positrons | | |
| Protons | | |
| Resonances | | |

Figure 1: Presenting knowledge for browsing

If we use words for indexing the only way we can mechanically display the subject index of a collection is alphabetically. Two concepts will find themselves in close proximity not because of their semantic similarity but rather owing to the accidence of their names. Classification on the other hand groups concepts semantically, according to the closeness of their meaning. But in order to 'fix' this systematic organization, classification needs a notational device: a numeric, alphabetic or alphanumerical symbol that represent the class arrangement and supports its mechanical manipulation. Notation is the shortest possible way of expressing sometimes a very complex subject and is very useful in labelling physical documents or metadata for the purpose of systematic arrangement. In Figure 1 we see an example of a numerical, decimal (fractional) notation, where each digit represents a decimal level that corresponds to the level of subdivision in which the dot after every third digit is inserted only for the convenience of reading. We call this type of notation hierarchically expressive: the longer the notation the more specific the class it represents, and by removing the last digit we automatically broaden the class. Not all notations are decimal or hierarchically expressive; they can be a simple ordering device as is the case with LCC and BC2:

| | |
|---|---|
| Q | Science |
| QD1-999 | Chemistry |
| QD241-441 | Organic chemistry |

*Example from LCC*

| | |
|---|---|
| AZ | Science |
| C | Chemistry |
| CO | Organic |

*Examples from BC2*

Notations (classmarks) represent the class meaning in a universal way no matter how many terms we use to describe it, or in which language these terms may be. For example, if we use 536 to represent Heat (Thermodynamics) in the bibliographic services of China, Russia or the United Kingdom we will be able to search for documents on this topic, irrespective of the language or script in which they are published:

| DDC | | UDC | | |
|---|---|---|---|---|
| 536 | Heat | 536 | Heat. Thermodynamics | [English] |
| | Chaleur | | Chaleur. Thermodynamique | [French] |
| | Теплота | | Тепло. Термодинамика | [Russian] |

*Examples from UDC and DDC summaries*

Precisely because classification uses notation for indexing, which is practical for human communication, word access to schedules is of extreme importance. Notation descriptions, technically called captions, are one source of language terms, but they are to a large extent only indicative and are designed for the purpose of schedule presentation. Therefore, all widely used classification systems have carefully constructed and detailed subject-alphabetical indexes. In addition to this, databases for the management of classification will usually be designed to generate a caption hierarchy and/or chain indexes (contextualised terms within the hierarchy). Many online systems will provide controlled keywords attached to each class as well as mappings to subject heading systems if appropriate. The importance of word access to classification codes is such that alphabetical indexes to classification have become sophisticated terminology tools in their own right. For instance, DDC's Relative index, which relates concepts and terms scattered across the scheme, is widely accepted as one of the finest examples. UDC editions in different languages are known to be published with various kinds of indexes from a relative chain index, to indexes in the form of a thesaurus. S. R. Ranganathan, the author of CC, particularly favoured an alphabetical index in the form of a chain index.

## 2.1 The role of logical and perspective hierarchies

The most prominent characteristic of all classifications is hierarchy: the organization of concepts into a sequence of classes based on various principles of division, following the order of classes from broader to narrower. Hierarchy is important as it determines the meaning of a concept which operates above the level of language. For example, the relationship between the concept of *carrot* and the concept of a *root vegetable* is always the same and universally understood, irrespective of which words in which language we use to name these entities. However, when classifications present an entire universe of knowledge, they need to deploy several different structural and organizational mechanisms, logical hierarchy being only one of them.

### 2.1.1 Logical hierarchy

The general principle of a valid hierarchy is that all members of a class share at least one characteristic. Each class contains a number of subclasses and all members of a subclass are contained in the class above and inherit its characteristics. These are the principles of a logical hierarchy that may be built on three kinds of hierarchical relationships: generic or proper hierarchic relationships (genera-species where each member is a *kind of* the class above); partitive relationships (whole-part where each member is part of the class above), member class and instance relationships (each member is an instantiation of a class or simply its member based on an arbitrary characteristic). The most typical characteristic of a logical hierarchy is the fact that each entity will have only one place in the hierarchy. Every entity will have one broader class, the property of subdivision will be transitive and the logic of the hierarchy will support broadening, narrowing or exploding searches in the process of information retrieval.

For instance, a logical hierarchy of languages can help in avoiding confusion between the term *Tepehua* , one of Totonacan languages and *Tepehuan*, a Uto-Aztecan language. While it is easy to confuse the names (Tepehua / Tepehuan), the hierarchy and its notational representation indicate that these are two entirely different concepts:

| | |
|---|---|
| = 82 | Indigenous languages of western North American Coast, Mexico and Yucatan |
| ... | |
| =821 | Penutian, Huave, Utian, Totonacan, Mixe-Zoque languages |
| =821.22 | Totonacan languages |
| =821.221 | **Tepehua** |
| ... | |
| =822 | Uto-Aztecan and Kiowa-Tanoan languages |
| =822.2 | Uto-Aztecan languages |
| =822.24 | Southern Uto-Aztecan languages |
| =822.248 | **Tepehuan** |

*Example from UDC MRF (2010)*

Hierarchy is even more important in cases of polysemy when the meaning of a word can only become evident through its contextualisation within a hierarchy. For instance, the meaning of a polyseme such as *cell* will depend on whether it is placed within the broader class of *cytology*, subsumed to the concept of *prison* or to the concept of *electrochemical equipment*. The need for disambiguation increases with the size of the knowledge area covered and becomes very important in searching large collections comprising numerous forms of knowledge.

### 2.1.2 Perspective or aspect hierarchies

General bibliographic classifications arrange the universe of knowledge by placing different forms of knowledge into main classes (macro level) according to some existing, widely accepted theories of knowledge organization. This is determined by the basic function of knowledge mediation which is facilitating access to knowledge and its use: phenomena and topics that are studied and sought together are placed together. By respecting this principle, topics for, e.g. sport fishermen such as kinds of fishing, fishing by different methods and tools, fishing by type of fish, by place, by time, etc. will be collocated together. As soon as we start organizing knowledge in this way we will find that many phenomena will be of interest to many fields of knowledge: gold will be studied as a chemical element in chemistry, as a product in the mining industry or as a material in arts and crafts. This kind of approach to knowledge organization in which phenomena are subsumed to the form of knowledge within which they are studied is known as a disciplinary knowledge organization.

Concepts organized in main classes corresponding to disciplines and sub disciplines will therefore be related not only through a logical hierarchy (subordination and super ordination) but may find themselves in collateral relationships since the same concepts will appear in numerous other hierarchies within the scheme (Bhattacharyya, 1980). This will result in perspective or aspect hierarchies (Svenonius, 1997). Figure 2 shows the occurrence of the concept of shark in different knowledge fields:

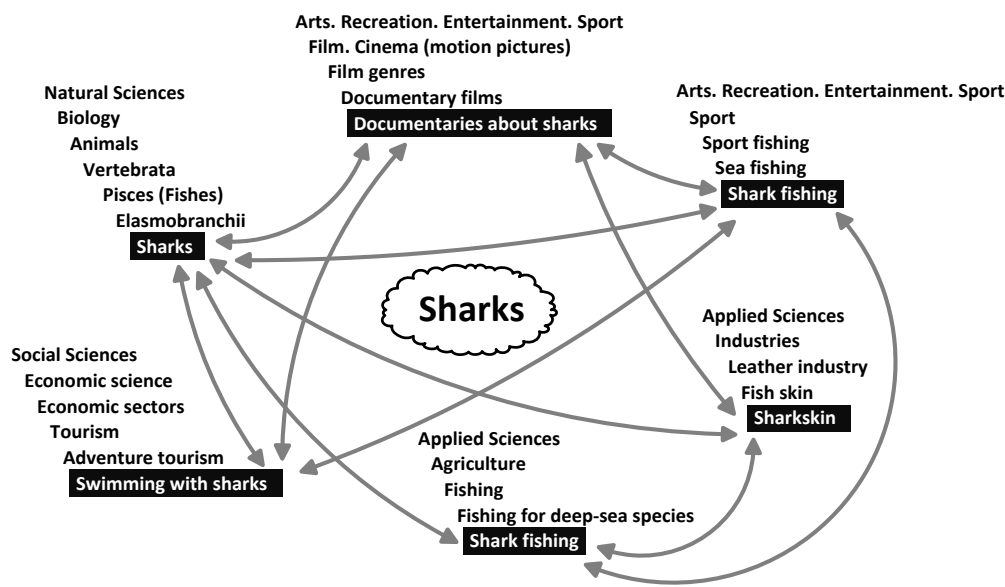

**Figure 2:** 'Distributed relatives', one concept in different fields of knowledge

A single logical hierarchy will support broadening, narrowing or exploding searches within the same subdivision. Perspective hierarchies are a standard feature of universal bibliographic classifications that have an important impact on information retrieval as they can support concept association, contextualisation and disambiguation across different fields of knowledge. The fact that the same concept can appear in different hierarchies leads to four important structural features typical of bibliographic classifications:

**a) Syndetic structure** based on the large number of associative relationships (see also references) that are established between classes across the entire scheme:

| 176 Sexual ethics. Sexual morality | KNOWLEDGE FIELD |
|---|---|
| → 173  Family ethics | [Philosophy] |
| → 2-447  Sexual relations, activity | [Religion] |
| → 316.36  Marriage and family | [Sociology] |
| → 343.5  Offences against public credit, morality, the family | [Law] |
| → 351.764  Supervision of sexual morality. Control of prostitution. Supervision of brothels | [Public Administration] |
| → 364.633  Sexual abuse | [Social Welfare] |
| → 392.53  Abstinence from matrimony or sexual relations | [Cultural Anthropology] |
| → 613.88 Sexual hygiene. Sexual education. Sex life | [Medicine] |

*Example of a class and its see also relationship from UDC*

Not all bibliographic classifications have the same affinity towards connecting distributed relatives. This is usually a feature of classifications that are created to function as a coherent system (e.g. DDC, UDC). LCC and BC2 are developed as series of special classifications and associative relationships between different fields of knowledge are significantly less frequent and in many cases non-existent.

**b) Relative subject alphabetical index** constitutes a standard part of a classification system and serves to connect 'distributed relatives' and support retrieval and contextualisation of search, e.g.

| **Marriage** | **306.81** |
|---|---|
| arts | 700.454.3 |
| customs | 392.5 |
| ethics | 173 |
| religion | 205.63 |
| ... | |
| folklore | 398.27 |
| law | 346.016 |

*Example of Relative Index (DDC)*

**c) Separation of common concepts** (isolates) from classification of knowledge. Owing to their frequency, general concepts are organized as separate facet hierarchies or tables, usually called common auxiliaries, so that they can be reused and combined with concepts in knowledge fields. This approach can be seen in DDC, UDC, CC, BC2 and many other general knowledge schemes

| Place facet | | Combinations with Place | |
|---|---|---|---|
| (4) | Europe | 027 | General libraries |
| (46) | Countries of the Iberian Peninsula | 027(469) | General libraries in Portugal |
| (469) | Portugal | | |
| | | 338.48 | Tourism |
| | | 338.48(469) | Tourism in Portugal |
| | | 726 | Church buildings |
| | | 726(469) | Church buildings in Portugal |
| | | 91 | Regional geography |
| | | 91(469) | Regional geography of Portugal |
| | | 94 | History |
| | | 94(469) | History of Portugal |

*Example of common concepts in use (UDC)*

**d) Parallel divisions**, are a typical feature of many bibliographic classifications and will occur frequently between disciplines that deal with the same phenomena: Natural sciences and Applied sciences; Languages, Linguistics and Literatures; Geography and History etc. For instance, the classification of plants in Botany or animals in Zoology will be used to build hierarchies in Palaeobotany or in Agriculture. In the example from DDC below we can see how a concept hierarchy of Insects from class 590 Animals is not replicated in class 560 Paleozoology. Instead there is an instruction explaining that fossils of insects should be built using numbers from the class hierarchy at 590 when required:

| | | | |
|---|---|---|---|
| 560 | Paleontology. Paleozoology | 590 | Animals |
| 565 | Fossil Arthropoda | 595 | Arthropoda |
| **565.7** | **Insecta** | **595.7** | **Insecta** |
| | *Add to base number 565.7* | 595.72 | Apterygota |
| | *the number following 595.72-595.79* | 595.73 | Exopterigota (Hemimetabola) |
| | *e.g. Coleoptera 565.76* | 595.74 | Mecoptera, Trichoptera, Neuroptera... |
| 566 | Fossil Chordata | ... | |
| ... | | 595.79 | Hymenoptera |

*Example from DDC (22)*

## 2.2 Analytico-synthetic principle in building classifications

The number of knowledge forms which are selected as a macro structure of knowledge (main classes) and the logic of their sequence in bibliographic classifications is usually based on some widely accepted philosophical knowledge classifications which are themselves subject to many theoretical discussions and criticism. It is well known that any such disciplinary classification is inevitably rigid with respect to the dynamic nature of knowledge and biased in terms of a cultural or philosophical point of view. No matter how frequently we change this view, it will always be a single, frozen snapshot of a world at a given time.

To address this problem, bibliographic classifications have developed various mechanisms, methods and techniques that operate at the 'micro level' of a classification and help in resolving intra-, inter- and cross-disciplinary relationships as they appear in the subject of documents. This is resolved in three ways:

- identification of isolates or general concepts that are frequently reused across knowledge and their placement in separate facets. These would usually be isolates of place, time, languages, forms of documents, general characteristics of various kind (as explained in the previous section)

- organization of concepts within a knowledge field into a set of mutually exclusive broad concept categories that can be combined to express the composite nature of any given subject and all aspects from which a phenomenon may be studied and viewed within a single field of knowledge

- syntactical rules and devices to express relations between concepts from different disciplines.

Classification schemes created in such a way are called analytico-synthetic.

## 2.2.1. Facet analytical theory

If we take a document such as "Economic development in Europe after 1945", we recognize some general categories of elements: 'economy' is the main subject, that 'development' is a process, 'Europe' is a place, and 'after 1945' is a time. These categorizations are valid whether we analyse the content of the document or are trying to formulate our query when searching. It has been widely accepted in the theory of content analysis and indexing that subject topicality can be analysed and expressed with a high level of accuracy if we follow a particular set of fundamental concept categories (Hutchins, 1975; Langridge, 1989; Broughton, 2004).

Although a kind of categorical analysis was intuitively built into the UDC back in early 1900s as a way of structuring schedules, the categorical structural principle was not properly analysed and introduced to classification theory before the work by the work by the Indian classificationist S. R. Ranganathan in the 1930s. He proposed five fundamental categories of concepts or fundamental facet categories: personality, matter, energy, space and time as a categorical framework for building classification for all fields of knowledge and proved his theory by building his Colon Classification (CC), the first fully faceted bibliographic classification system. The principle of facet analysis in organization of knowledge was widely discussed and was adopted as efficient technique in subject organization from the 1960s onward under the name of *facet analytical theory*. The extended set of 13 fundamental facet categories (Thing-Kind-Part-Property-Material-Process-Operation-Patient-Product-Byproduct-Agent-Place-Time) was proposed by the Classification Research Group (CRG), the British group of classification experts in the 1960s (Hutchins, 1975). Taking this as a fundamental principle, CRG commenced their work on restructuring Bliss Bibliographic Classification into a new BC2 faceted scheme in the 1970s. The principle has been tested ever since and used in building thesauri and classifications.

Facet analytical theory as proposed and implemented in BC2 defines the principles not only for the analysis of a subject field and its categorical structuring in a bottom-up fashion but suggests what is defined as the correct order of the facets in the schedule (filial order) as well as the strict order of citing of categorical elements in a complex subject statement (citation order). Once concepts are hierarchically organized into their respective broader facet categories, their arrangement and combination is fixed and mechanised through an inexpressive notational system (Broughton, 2004):

| | |
|---|---|
| PBK | Sacred books |
| ... | |
| PIJ | Vedic religion |
| PIJ BK | Sacred books |
| PIJ C | Samhitas |
| PIJ D | Trayi Vidya |

*Example of notation from BC2*

In the example above the notation itself does not indicate that class PIJ C Samhitas is super-ordinated to class PIJ D Trayi Vidhya or that the notation of class PIJ BK Sacred books is composed of PBK Sacred books and PIJ Vedic religion. The notation in BC2 has primarily 'ordinal' value as the use of the classification is assumed to serve the purpose of shelf arrangement for which the brevity of notation is most important. Faceted classifications can also have a hierarchically and syntactically expressive notation (Broughton & Slavic, 2007), e.g.:

| [By Language] | | [By Literary form] | | [By Period] | | [By Document form] | |
|---|---|---|---|---|---|---|---|
| A | English literature | -1 | poetry | aa | old period | 01 | polygraphic work |
| B | Hindi Literature | -2 | drama | ab | middle period | 02 | collected works |
| C | Arabic Literature | -3 | prose | ac | modern period | 03 | selected works |
| ... | | ... | | ... | | 031 | anthologies |
| | | | | | | ... | |

| | |
|---|---|
| A-1aa031 | English literature - poetry - old period - anthologies |
| B-2ac02 | Hindi literature - drama - modern period - collected works |
| C-1ac03 | Arabic literature - poetry - modern period - selected works |
| C-3ac031 | Arabic literature - prose - modern period - anthologies |

*Model of syntactically expressive notation*

The first and most important step in building a classification structure of this kind is in deciding what constitutes the main, thing or entity facet for that particular knowledge field. Once this is decided the other broad facet categories are formed around the main facet. A decision of what is the main facet in, for example, Information Science (services/institutions or operations/functions or documents), or Medicine (organs, physiology, diseases, patients or treatments) is an arbitrary decision by classification designers and is based on their perception of what may be the most logical or most useful order of subject within a field of knowledge, for an assumed application.

The difference in the number of fundamental categories between CC and BC2 clearly indicates that there are some issues concerning the very 'fundamental' nature of these categories (Hjørland, 2006). The universality of the principle can also be questioned with respect to the fact that CRG facet categories work well in natural and applied sciences but are not as easy to apply in social sciences and humanities (Svenonius, 1997). In addition, the explanation of this principle in building a classification 'bottom up', based on literary warrant seems to have little practical merit since, in reality, the terminology for building a classification has to be aligned with user warrant leading to an actual alignment of two different views: those of users and those of experts (Giess, Wild, McMahon, 2008).

## 2.2.2 Other approaches: facet analysis and free faceted classification

Even though this does not entirely agree with the facet analytical theory, organizing concepts into mutually exclusive facets is viewed as a flexible approach whenever one wants to manipulate, coordinate and re-order facet hierarchies to achieve different arrangements and emphasis. Such is the idea of 'freely faceted classification' as proposed by Gnoli & Mei (2006). Based on this is the proposal for free facets to be combined with a classification of phenomena, in contrast to disciplinary classifications. In such a scheme knowledge phenomena are not subsumed to disciplines but vice versa, following the tradition of Subject Classification by J. Brown, which was a classification of phenomena used at the beginning of the 20th century in a number of American and British libraries (Szostak & Gnoli, 2008)

Although it has little to do with the principles of facet analytical theory, the advantage of organizing entities, objects or concepts into facets to improve hierarchy management, display and browsing has also become apparent to the Web designers. This kind of vocabulary organization is widely accepted as good practice outside the specific field of subject analysis and knowledge classification and popularised in the browsing interfaces of portals, commercial websites and content management systems (La Barre, 2006; Vickery, 2008).

## 2.2.3. Relating subjects from different fields of knowledge or phase relationships

One of the most prominent issues in subject indexing arises from the fact that documents may represent and relate subjects from different fields of knowledge: "The use of granite in altar-domes in Mediaeval Churches of Italy", "Scholarly production in bibliographies of European national libraries", "Application of computers in education", "Bibliography of Rare books". Synthesis based on facet analytical theory, as described in the previous section, helps resolving the situation in subject indexing when we can assume a single principle concept modified by a cluster of subsidiary concepts which can be analysed and presented as a compound notational expression. Classifications designed for the arrangement of library shelves will assume that, no matter how many subjects are discussed in a document, it will always be classified in place depending on which subject is deemed to be more appropriate for given content.

Classifications, designed for detailed indexing and retrieval usually have a set of notational symbols (relators) for establishing relationships between two or more independent or otherwise unrelated subjects. For instance, UDC uses + symbols, when subjects are connected and studied as juxtaposed, e.g. 73+75 Sculpture and Painting; a colon when subjects are studied in a simple relation e.g. 73:75 Relationship between Sculpture and Painting. These kinds of relationship are termed *phase relationships* and they have been thoroughly studied and widely discussed in the theory of classification and indexing. Jean Perreault offered a detailed schema of these relators based on the earlier work by J. E. L. Farradane, J. C. Gardin and E. Grolier (Hutchins, 1975).

## 2.3 Enumerative vs. analytico-synthetic classifications

Traditionally, classification theory makes a distinction between enumerative and analytico-synthetic classification schemes and this frequently causes confusion as many classifications combine these principles in different ways. Both approaches to structuring schedules have their advantages and their shortcomings when it comes to their application in information retrieval or information integration in an online environment.

Analytico-synthetic classifications allow the synthesis of concepts in the process of indexing and their coordination in the process of searching, i.e. parsing of composite subject expressions to their composite elements. This means that although subjects such as 'history of mathematics', 'wars in Africa in the 19th century', 'English poetry' or 'herbal medicine in melanoma therapy' do not appear in the schedules enumerated as such, they can be expressed. The obvious advantage of this approach is that such schedules are suited to covering subjects in an unlimited way, in great detail, with a smaller but logically structured vocabulary. The disadvantage is that when one scans these schedules in order to extract the vocabulary, the analysis produces only a very partial picture of the actual subject coverage. To address this issue, analytico-synthetic classifications try to enumerate combinations (BC2) and others try to go halfway by listing a selection of examples of combination (UDC).

Enumerative classifications are typically created for the arrangement of library shelves and their main concern are short classification codes (for both simple and complex subjects) that are easy to file and that take little space on book spines. The easiest way to achieve this is to publish a scheme with a selection of the subject combinations that are most likely to appear in an imaginary collection and assign to them simple, short, ordering notations. The disadvantage are twofold: limited power in indexing, as obviously one cannot plan for all combinations of subjects; and occasional compressions of subdivisions into a single class, resulting in illogical and awkward placement of compound subjects.

The example below shows how DDC enumerates the subdivision of Organizations by place. Each of these composite classes is assigned a simple notation. In UDC when determination of place is required, the code of place is taken from the place facet, i.e. common auxiliaries of place, and combined with the main subject notation:

| **Enumerative classification (DDC)** | | **Analytico-synthetic classification (UDC)** | |
|---|---|---|---|
| 060 | General organizations & museum science | 06 | Organizations of a general nature |
| 061 | Organizations in North America | | *Examples of combination:* |
| 062 | Organizations in British Isles; in England | 06(7) | Organizations - North America |
| 063 | Organizations in central Europe; in Germany | 06(41) | Organizations - British Isles |
| 064 | Organizations in France & Monaco | 06(430) | Organizations - Germany |
| ... | | ... | |
| | | | |
| 070 | News media, journalism & publishing | 070 | Newspapers. The Press. Journalism |
| 071 | Newspapers in North America | | *Examples of combination:* |
| 072 | Newspapers in British Isles; in England | 070(7) | Newspapers - North America |
| 073 | Newspapers in central Europe; in Germany | 070(41) | Newspapers - British Isles |
| 074 | Newspapers in France & Monaco | 070(430) | Newspapers - Germany |
| ... | | ... | |

*Example from DDC and UDC Summaries*

From the above example it is clear why enumerative classifications require larger schedules to achieve the same indexing specificity hence LCC has over 200,000 and significantly weaker indexing power than UDC which has only 68,000 classes. Care should be taken not to generalise and to assume that 'enumerative' classifications such as DDC or LCC could not use synthesis and allow the building of numbers to a certain extent. Equally, analytico-synthetic classifications can include in their schedules pre-built notations.

### 3. Classification in information retrieval

From what has been said so far about logical and aspect hierarchies and alphabetical indexes attached to classification, we would expect that any information system holding classification data supports the following two simple term-search scenarios: a) a systematic display of results allowing for semantic search expansion as shown on the left illustration below; b) a systematic display of results enabling disambiguation and contextualisation of search terms within perspective hierarchies, in the right illustration below.

| | | |
|---|---|---|
| **hadrons** | | Search |

| | | Hits |
|---|---|---|
| 539.12 | Elementary Particles | 132 |
| 539.125/.126 | **Hadrons.** Baryons and mesons | 58 |
| 539.125 | Nucleons | 38 |
| 539.125.4 | Protons | 5 |
| 539.125.46 | Antiprotons | 2 |
| 539.125.5 | Neutrons | 7 |
| 539.125.56 | Antineutrons | 1 |
| 539.126.3 | Mesons | 9 |
| 539.126.5 | Resonances | 11 |
| 539.126.6 | Hyperons | 6 |
| **rabbit** | | Search |

| | | Hits |
|---|---|---|
| 569.32 | Zoology: Rodentia and Lagomorpha | 7 |
| 632.935.7 | Protection of Crops | 3 |
| 636.92 | Animal Husbandry: Domestic Rabbits | 38 |
| 636.92.045 | Animal Husbandry: Domestic Rabbits. Pets | 10 |
| 636.932 | Animal Husbandry: Rodents kept for fur | 9 |
| 639.112 | Hunting: Small game generally | 22 |
| 641.8 | Cooking: Main dishes | 2 |
| 677.534 | Textile industry: Hare fur. Rabbit fur | 8 |

**Figure 3**: Classification supporting simple term search

The most obvious place to look for good practice in using classification data to support subject searching, as illustrated above, would be library OPACs. Paradoxically this is not the case and this fact has been puzzling researchers for decades. Studies have so far confirmed very poor use of classification in spite of obvious advantages in its use (Wajenberg, 1983; Svenonius, 1983; Cochrane & Markey, 1985; Markey, 1986, 1990, 1996; Markey & Weller, 1996; Hildreth, 1991; Hancock, 1987). At the same time OPACs surveys indicate that users experience problems in finding the correct search term, increasing recall when results are too few and increasing precision when too many items are found (Larson, 1991; Yu & Young, 2004). In spite of the significant advances in database and interface technology, recent research of subject access functionality conducted on the same library systems in 2003 and 2008 in Italy, shows very poor general performance of most of the major vendor systems and almost no change over a 5 year period (Casson, Fabrizzi & Slavic, 2008).

The problems encountered when searching classification in library systems reside in a clear discrepancy between the composite structure and dense classification notation semantics, and their primitive processing in library systems. Back in the 1980s Wajenberg (1983) recommended extending MARC codes for DDC so as to use shelfmarks for improving subject searching of bibliographic data. Cochrane and Markey (1985) provided an exhaustive list of classification data elements needed to support searching and browsing of DDC. This effectively represented a subject authority control framework that was independent of the bibliographic description and was capable of serving classification maintenance, distribution and its subsequent use in information retrieval.

Although, by the 1990s, librarians fully embraced authority control for managing names, it took some time for the same model of the central management of vocabulary to be proposed as a solution for supporting subject data, including classification. Mandel (1995) emphasised the obvious fact that access points to classification data ought to be controlled centrally and independently of bibliographic records. Subject authority control significantly changes the way classification is displayed, browsed and searched; it allows independent access to notation, notation hierarchy, associative relationships and word access to classification. Figure 4 shows the way subject access points based on classification are managed in three languages, in the authority file of the Network of Libraries and Information Centers in Switzerland (NEBIS), www.nebis.ch.

**Figure 4:** Subject authority record, OPAC display

The first breakthrough with respect to classification use in library systems was in 1992 when the USMARC Format for Classification Data was created as an independent authority format. The creation of this format was driven by the plans for the conversion of LCC, to which purpose it was effectively used in 1993 (Guenther, 1996). This standard was later superseded by MARC 21 Concise Format for Classification Data which has since been updated and improved through subsequent versions. Although the format was, theoretically, presented as a general solution, in practice it was based on LCC and DDC structures and is for this reason better suited to enumerative classifications. The next opportunity to 'rethink' and improve machine readability was occasioned by work on the UNIMARC authority format for classification data that was meant to be improved to support analytico-synthetic systems. This standard is still awaiting its official completion (Slavic & Cordeiro, 2004; Slavic, 2008).

In parallel to the above mentioned automation of library systems, the publishers and owners of DDC and UDC created more advanced database tools, primarily to facilitate the maintenance and publishing of classifications. The issue of classification data being able to support user friendly hierarchical browsing, semantic linking and facet control became central to classification owners when publishers started selling access to schedules on CD ROM (from 1995), and on the Web (from 2001). The contribution to the automation of classification from the publishers of classification data is significant, for when a classification is converted into a machine readable format, the same data can be used to support various tools for machine assisted indexing or authority control tools for supporting information retrieval (Markey, 2006). As more and more classification data are distributed to users as files that can be ingested in information systems, the concern over the inability of bibliographic standards to handle these data is becoming greater. Fortunately, the development of formats and standards for sharing not only classification but all other KOSs has has exceeded the limits of specific domains and has been an active area of W3C development over the past ten years.

## 4. Bibliographic Classifications on the Web 1990-2010

### 4.1 Subject gateways (SG), hubs and portals

The idea that bibliographic classification can be used to 'organize the Web' was very appealing to Web developers at the beginning of the 1990s, when the Internet was predominately academic, cultural and research-orientated. This reflected a strong commitment to interoperability and standardization in information sharing. Web development in the mid-nineties was marked by numerous research projects building quality SGs. This trend was especially strong in the UK where a number of SGs were developed within the Electronic Libraries Programme (eLib). In 1999 some of these services joined the Resource Discovery Network (RDN) consisting of subject hubs such as SOSIG (social sciences hub), EMC (engineering, maths, computing hub) and Humbul (humanities hub). This led to a further federation of services that shared the same harvesting, metadata population and indexing tools which eventually evolved into a hub called Intute (http://www.intute.ac.uk) in 2006, a catalogue of quality selected Internet resources built from the records of eight subject services (Kerr, 2009).

In the beginning, the main characteristic of these quality SGs was that they all used UDC and DDC outlines as the basic browsing structure. As services expanded to include more subjects and more subject specific indexing languages, and as the number of resources requiring description expanded with the growth of the Web, SGs ceased using UDC or DDC for the subject browsing interface. This was a logical consequence of the fact that services needed to provide browsing shortcuts to subject areas, some of which were on the fifth or sixth levels of general classification subdivision and as such generated an unnecessary long sequence of browsing steps. Classification was, however, retained for content indexing by the system 'behind the scenes' to support consistency in indexing independently from the browsing function (Slavic, 2006).

## 4.2 Automatic classification of web resources

An important aspect of the use of classification in subject gateways (SG) was that it led to an exploration of automatic classification of textual web resources using bibliographic classifications. GERHARD - German Harvest Automated Retrieval and Directory (1997-2002) was one such research project that created a gateway of academic resources on the German web. GERHARD was a database-driven robot that collected academically relevant documents which were indexed using computer-linguistic and statistical methods and classified by UDC. GERHARD's architecture consisted of a database-driven gatherer, automatic classification and an integrated searching and browsing service. The generated metadata and index of documents were held in a database that towards the end of the project contained 1,300,000 records. Automatic classification was based on the UDC trilingual authority files from the ETH library in Zürich containing around 70,000 classes. This authority file supported searching of compound and complex UDC numbers and around 15 different relationships that can be established between individual UDC numbers (Möller et al., 1999).The gateway was taken down in 2006 following a shortage of funding.

In the USA, OCLC's research on the potentials of DDC use on the Web was linked to a number of other projects on cataloguing of Web resources that started in the early 1990s. *Scorpion* was one such project whose aim was to build tools for automatic subject assignment, combining automatic indexing techniques and DDC and was envisaged as an aid to human cataloguing by automating subject assignment where items are available electronically. Scorpion was also used as an automatic classification tool in Renardus, a gateway to the cultural and scientific collections on the European Web. The result of the project was an SG based on mapping different vocabularies and DDC with cross subject-gateway browsing enabled through DDC (Koch, Neuroth & Day, 2003). The Renardus gateway was also taken down when funding ceased.

Similarly to SGs, digital repositories and open archives (learning materials, electronic journals, research papers and theses) which began appearing in greater numbers after 2000 also use bibliographic classifications. Committed to scholarly communication, by their very nature, repositories usually opt for bibliographic classifications that are widely used in national or international bibliographic services since the interoperability of subject access is an important requirement for their integration into national information networks (Koch, 2006).

The evolution of the Web, from purely academic to commercial, business and social coincided with a significant improvement in technology. Web portal developers started to be primarily interested in the benefits of concept organization into practical and purposeful categories that their users/customers could easily combine or independently navigate. Subject orientated, simple hierarchical structures on gateways and portals have started to be combined with, or replaced by, a faceted organization of object/subject properties and attributes. These applications led to the development of data formats and tools for the management of faceted vocabularies and their use in a Web interface (La Barre, 2006; Giess, Wild & McMahon, 2008).

## 4.3 Classification as a pivot and terminology services

The idea of using classification as a 'switching language' or 'pivot' to map indexing languages for the purpose of information integration and exchange was widely discussed in the 1970s (Marcella & Newton, 1995). This is even more relevant today with the overall trend of information integration on the web. Classifications are particularly well suited for use as a central mapping spine in connecting the vocabularies of different languages and seem to give visible results quickly and easily.

In 1999 OCLC started the CORC project to provide tools for the co-operative creation, maintenance and use of metadata for Web resources which uses previous project material such as a tool for automatic classification developed in *Scorpion*. The goal of the project was to support access control for names or classification numbers in cases where different content can be supplied for different types of user (e.g. the meaning of a classification number could be given in different languages, or related to different user ages or education levels). CORC also explores XML/RDF architecture in order to enable access and make use of existing authority files rather than creating new authority records. CORC records link DDC numbers with LCSH and this linking gathers a number of related subject headings under one classification number (Hickey & Vizine-Goetz, 1999). Since 2003, OCLC developments have been moving towards terminological services that provide shareable vocabulary data using open vocabulary encoding standards, XML and Web services. Within these initiatives there is a number of significant developments in vocabulary mapping, e.g. DDC and LCSH, LCC, National Library of Medicine Classification (NLM) and DDC (Vizine-Goetz et al., 2004).

In Europe, examples of classification use in such context are also numerous. Concordances between DDC and UDC were prepared for the Czech Uniform Information Gateway in order to make it interoperable with gateways using DDC. The MSAC (Multilingual Subject Access to Catalogues of National Libraries) project focused on mapping subject heading systems in seven central European countries through UDC (Balikova, 2005). HILT (High-Level Thesaurus) was a UK project focused on linking subject vocabularies across a range of communities and services to support subject cross-searching and browsing. In its latest phase in 2009, it provided facilities for searching 11 subject vocabularies linked through a DDC outline, and m2m pilot testing on several SGs. At the moment The European Library and Europeana projects are focusing on interoperability and mapping between vocabularies.

## 5. Classification for computers
According to Soergel (1999: pp 1119) classification can have many functions in information retrieval: providing semantic road maps; improving communication and learning; providing a conceptual base for the design of research; providing classification for actions; supporting information retrieval; providing a conceptual basis for knowledge-based systems; providing the conceptual basis for data element definition and object hierarchies in software systems; cross-discipline, cross-language and cross-culture mapping; and serving as a base for natural language processing. This list of applications implies a strong relationship between classification and computers.

From the overview of classification use in library systems and on the Web it is evident that, in spite of the fact that the intellectual content of a classification may be useful, there is very little use for it in information retrieval unless its content is machine readable and distributed in some kind of standardized format that can be automatically ingested and processed by information systems. From the use of classification on the Web, however, it is obvious that domain specific standards such as MARC are of little value once classification starts to be shared outside a specific domain. Standards are vehicles for KOS implementation and sharing as recent developments can illustrate. For instance, ISO/IEC 13250 Topic Maps, BS 8723 Structured vocabularies for information retrieval and Simple Knowledge Organization System (SKOS), have all been created to support sharing and use of subject vocabularies across systems and domains. The general idea behind these and other such standards is that the availability and exchange of controlled vocabularies of various kinds, and subject indexing languages in an open networked environment may contribute to resource discovery through referencing, resolving language ambiguities and providing semantic context for text processing. With vocabularies expressed in a standardized way, there is a realistic prospect of centrally managed vocabulary repositories and services that will facilitate cross-collection resource discovery through vocabulary mappings and translation (Tudhope, Koch & Heery, 2006).

Through the use of an XML/RDF encoding schema that allows unique identification of data elements via a URI, and their automatic linking and referencing on the open Web, standards such as SKOS enable a more advanced way of processing semantics. Furthermore, indexing languages published in this way can then be combined and extended with various machine-understandable statements or inference rules using a formal web ontology language such as OWL (Web Ontology Language). Such an approach can lead to m2m processing of semantics captured in indexing languages and is considered one of the building blocks of the SW. With the advancement of XML/RDF technology, and the wider application of SKOS in publishing and linking subject vocabularies and document metadata, we can see how part of the SW scenario is unfolding. Currently, thesauri, classifications and collection metadata are released as SKOS data and are open for m2m processing and linking. The more data is published in this way, the greater the chance and incidence it has of being meaningfully and usefully connected. For example, connecting geodata to Wikipedia, to library collections, to archival or museum collections looks very appealing from the point of view of research and academic community.

However, for this scenario to work, both library collection metadata and KOSs ought to be published on the Web as free and open for automatic processing and semantic linking. The publishing of library catalogues as linked data is supported through initiatives such as the W3C Library Linked Data Incubator Group (LLD XG). The publishing of bibliographic classifications schemes as linked data using SKOS XML/RDF is confronted with two obstacles. The first is that SKOS data encoding model itself is not entirely suited to encoding the richness of semantic relationships and the structural complexity of universal classification, as we described in Section 3. The second, and probably more problematic, is related to the issue of copyright and restrictions imposed by publishers and owners of classification systems.

Brickley (2009) suggests that hybrid approaches may be the correct way to proceed: human classification mixed with automatic classification, professional subject schemes mixed with semi-structured user tagging, thing-oriented ontologies mixed with perspective and discipline-oriented subject classifications. In his opinion the most pressing issue is that of sharing vocabularies on the open Web: "By sharing what we know on the Web, by publishing every major classification system, thesaurus and subject headings system as a natural and essential part of the public information record, we open the door to the kind of hybrid models outlined above. There are all kinds of obstacles: cultural, technical, legal, practical. But the goal is inescapable. Subject classification systems and thesaurus systems which are not widely available in open formats will lose ground to those which are."

## 6. Concluding remarks

We have established that classifications are widely used in the bibliographic world, in national bibliographies and library collections. From a more detailed analysis of some features of these systems it is evident that they have a very comprehensive vocabulary, structured and semantically related in the most intricate way which is supported by natural-language terms in the form of subject-alphabetical indices. National SGs and portals organizing resources based on bibliographic classification have been continuously appearing since 2000, particularly in Central and Eastern Europe. As SGs evolved from simple directories to subject-orientated hub services, they started to use classification in a more sophisticated way: as a source of vocabulary for automatic classification and finally as a method of controlling and linking subject vocabularies behind the system. In this context subject vocabularies started to be managed as authority tools and shared between metadata repositories in which classification was used as a means of semantic control. As SGs are established as academic research projects with no long-term funding or adequate business model, similar to many other Web services, they tend to have a short lifespan, but it is not easy to predict when and if bibliographic classifications will cease to be used in this way.

In spite of the intensive use of classifications on the web there are relatively few studies measuring or assessing their usefulness in supporting information browsing. Vickery (2008: 3) concluded that online retrieval on the Internet has "largely done away with the need to mechanise sequential spatial display, but that the search result display could be considerably improved if systematic browsing options were provided". Soergel (2009) reminds us that we have never actually seen a proper implementation of knowledge classification in a modern information retrieval system. One can only hope that such solutions will emerge with the wider use of classification schemes online. The increasing trend of sharing of KOSs on the web will influence the way we use classification systems in resource discovery in the future: hopefully as a source of complex concept relationships and an intricate interplay of subjects on which it would be possible to compute and mine for information and knowledge.